# Ferrimagnetic nanostructures for magnetic memory bits

*AUTHOR NAMES*

A. A. Ünal,[1] S. Valencia,[1] D. Marchenko,[1] K. J. Merazzo,[2] F. Radu,[1] M. Vázquez,[2] and J. Sánchez-Barriga[1,*]

AUTHOR ADDRESS

[1]Helmholtz-Zentrum Berlin für Materialien und Energie, Albert-Einstein-Str. 15, 12489 Berlin, Germany

[2]Instituto de Ciencia de Materiales de Madrid, CSIC, 28049 Madrid, Spain

**ABSTRACT**

Increasing the magnetic data recording density requires reducing the size of the individual memory elements of a recording layer as well as employing magnetic materials with temperature-dependent functionalities. Therefore, it is predicted that the near future of magnetic data storage technology involves a combination of energy-assisted recording on nanometer-scale magnetic media. We present the potential of heat-assisted magnetic recording on a patterned sample; a ferrimagnetic alloy composed of a rare earth and a transition metal, $DyCo_5$, which is grown on a hexagonal-ordered nanohole array membrane. The magnetization of the antidot array sample is out-of-plane oriented at room temperature and rotates towards in-plane upon heating above its spin-reorientation temperature ($T_R$) of ~350 K, just above room temperature. Upon cooling back to room temperature (below $T_R$), we observe a well-defined and unexpected in-plane magnetic domain configuration modulating with ~45 nm. We discuss the underlying mechanisms giving rise to this behavior by comparing the magnetic properties of the patterned sample with the ones of its extended thin film counterpart. Our results pave the way for novel applications of ferrimagnetic antidot arrays of superior functionality in magnetic nano-devices near room temperature.

**INTRODUCTION**

The research on artificially engineered spintronic materials in the form of layered thin films and low-dimensional structures has experienced a tremendous boost due to the versatility of their applications in computing technology. This is reflected by the development of magnetic tunneling based read-out technologies [1], magnetic sensors [2,3], and non-volatile magnetoresistive random-access memory devices [4]. Because conventional magnetic data storage relies on two-dimensional arrays of magnetic bits, developing higher density, cheaper and faster devices relies on reducing the lateral size of individual memory elements or data storage bits [5]. Ultimately controlling the magnetic anisotropy and shape of nanostructures, as well as understanding their magnetic properties in reduced dimensions are key aspects that have received a sustained interest over the past decades [6]. For instance, in antiferromagnetic (AF) materials, which have crucial importance in exchange-bias systems, finite-size effects lead to a scaling of the magnetic properties such as the ordering temperature or the magnetic anisotropy [7], while in ferromagnetic (FM) materials control of the shape anisotropy can be achieved depending on the geometry of the objects [8].

Nanostructured magnetic materials often exhibit novel properties over their bulk counterparts as a consequence of reduced atomic coordinates and modified density of states. Magnetic systems of low dimensionality like artificially grown or self-organized FM nanostructures [9,10], arrays of nanostructures [11,12], nanowires [13,14,15], and antidot arrays [16] have been intensively investigated, mainly triggered by the discovery of spontaneous self-organization of magnetic FePt nanoparticles on a surface [11]. In all these fields, there is a rapid decrease of the relevant length scales of the magnetic structures down to the sub-100 nm range. To illustrate this, a storage density of 1 Tbits/in$^2$ would require a bit length of around 25 nm. The underlying idea is to replace the comparatively large randomly-oriented magnetic grains in conventional media with a nanoscopic region of a single magnetic domain [17]. As the magnetic stability of individual particles scales with the material anisotropy constant and the particle volume, the use of FM FePt alloyed nanoparticles with very high magnetic anisotropy (only in their chemically ordered face-centered tetragonal L1$_0$ phase) has been considered one of the promising routes towards future ultrahigh-density recording-media applications [18,19]. In this case, the suppression of superparamagnetic effects at very small volumes might prove crucial to achieve stable and enhanced magnetic properties at the smallest length scales. On the other hand, the use of ferrimagnetic nanoparticles with controlled coercivity near room temperature might be a more desirable compromise to additionally achieve full control over the magnetic properties.

Recently, ferrimagnetic (FI) alloy systems with relatively low coercive fields, such as the cobalt (Co) - dysprosium (Dy) alloy $DyCo_5$, have been proposed as ideal candidates for magnetic data storage and spintronic applications [20] owing to their flexibility in controlling a tunable perpendicular exchange-bias effect at room temperature [21]. Here the Dy-Co alloy acts as a hard FI material with relatively low coercive field and well-defined perpendicular magnetic anisotropy at room temperature. The system exhibits a spin-reorientation temperature ($T_R$) of ~350 K above which the uniaxial magnetic anisotropy rotates from out-of-plane to in-plane. Such a remarkable change of the anisotropy axis within a narrow temperature window near room temperature renders this type of FI alloys one of the technologically relevant and promising materials for future spintronic applications. Therefore, particularly taking into account that the magnetic properties of $DyCo_5$ nanostructures has remained unexplored, patterning of this type of FI alloys down to the nanoscale as well as understanding their magnetic properties at ultimate length scales are both of scientific and technological interest.

Among the different growth methods of magnetic nanostructures, the anodization of alumina templates and the subsequent deposition of magnetic antidot arrays have certain advantages that overcome the unwanted effects of superparamagnetism. This technique employs ultrahigh-density substrates that are patterned down to the nanoscale as templates to grow thin films on top [16,22]. In the case of nanoporous alumina membranes, nanoholes are arranged in a fully-controllable hexagonal symmetry [23,24], and growing a thin film on top results in an array of antidots [25]. The antidots act as well-ordered nonmagnetic inclusions within the magnetic thin film, promoting the creation of single nano-domain structures whose sizes can be manipulated depending on the anodization conditions. This is a much less expensive technique compared to lithographic patterning methods. These nanostructures can be promising candidates for a new generation of ultrahigh-density magnetic-storage media mainly due to the absence of a superparamagnetic limit, as there are no isolated small magnetic entities. This is due to the fact that the nanoholes introduce a locally distributed shape anisotropy [26], act themselves as pinning centers for magnetic wall displacements, and their periodic distribution determines the whole magnetization process as well as the magnetoresistance behavior [16]. Investigations of FI antidot arrays with tunable and controllable magnetic anisotropies might in fact provide superior functionalities that have not been addressed so far.

In this work, we present a nanoscale-patterned, hexagonally-ordered FI $DyCo_5$ antidot array where the local nanoscopic distribution of the magnetization at room temperature can be controlled via heating the sample slightly above its spin-reorientation temperature. We report on controlling the magnetic anisotropy as a function of temperature and we compare the nano-patterned sample with its plain film counterpart. Using x-ray magnetic circular dichroism (XMCD) in combination with photoelectron emission microscopy (PEEM), we reveal that the temperature-dependent changes of the magnetic anisotropy in the presence of antidots results in well-defined and stable magnetic nano-domain configurations modulating within ~45 nm at room temperature. We discuss the underlying mechanisms giving rise to this behavior. Our results pave the way for future applications of FI nanostructures in heat-assisted magnetic nano-devices.

**METHODS**

We prepared antidot arrays of $DyCo_5$ by magnetron sputtering on top of hexagonally-ordered alumina membranes, which were fabricated using a two-step anodization process [20, 25]. For the present experiments, we produced membranes containing nanopores with a typical diameter of ~68 nm and center-to-center inter-pore distances of ~105 nm [see Fig. 1(a)]. A $DyCo_5$ extended film was grown on top of an un-patterned alumina substrate for comparison purposes. The growth of the antidot arrays and the extended film was performed simultaneously within the sputtering chamber, and their thickness was ~25 nm. The samples were obtained by co-deposition from Co and Dy targets keeping the substrate temperature at ~200°C. The co-deposition was done with a relative composition of 1:5 to ensure the spin-reorientation transition to be closest to room temperature. The samples were capped with a 2 nm thick Al layer to avoid oxidation upon transport in air. The magnetic properties of the extended films, their compensation and spin reorientation temperatures, coercive fields, and magnetic anisotropy axes were determined by element specific hysteresis loops at the Dy and Co absorption [21] edges using XMCD in transmission and scattering geometries [27].

High-resolution magnetic domain configurations around the nanoholes and across the spin-reorientation transition were obtained by XMCD-PEEM imaging [28]. Experiments were performed at the UE49-PGMa beamline of the synchrotron BESSY II. XMCD element-specific images were obtained by tuning the synchrotron photon energy to the $L_3$ and $M_5$ resonances of Co (779.2 eV) and Dy (1293.9 eV), respectively. Each of the XMCD images was calculated from a sequence of images taken with circular polarization (90% of circular photon polarization) and alternating the light helicity. After normalization to a bright-field image, the sequence was drift-corrected, and frames recorded at the same photon energy and polarization were averaged. The difference of the two resulting images with opposite helicity, divided by their sum, showed Co and Dy magnetic-domain contrasts, which represent the magnetization vector pointing along the incidence direction of the x-ray beam. Due to the 16º shallow incidence angle of the x-rays on the sample, our XMCD-PEEM measurements are mainly sensitive to the in-plane magnetization. The samples were priorly magnetized by applying a magnetic field of ~0.5 T perpendicular to the surface, so that the XMCD images in the initial remanent magnetic state at room temperature exhibited nearly-zero magnetic contrast.

**RESULTS AND DISCUSSION**

We investigated the formation of magnetic domains in FI DyCo$_5$ grown on extended substrates and on nanoporous membranes, controlling the sample temperature above and below the spin-reorientation transition temperature of ~350 K. Scanning electron microscopy (SEM) [Fig. 1(a)] and XPEEM images [Fig. 1(b)] from the nanohole array show the hexagonal arrangement of antidots. Circles and lines on the SEM image mark several punctual defects and structural domains that occur during the self-assembling anodization process. The distance between the borders of two nanoholes, where the metallic alloy is sitting, is around 37 nm. The XPEEM image shows photoelectrons emitted from the material deposited on the surroundings of the nanoholes as the bright intensity contrast, and dark contrast corresponds to the nanohole locations. By varying the x-ray photon energy and simultaneously recording XPEEM images, we obtain x-ray absorption spectra (XAS) for Co [Fig. 1(c)] and Dy [Fig. 1(d)], showing $L_{2,3}$ and $M_{4,5}$ absorption edges, respectively.

At a temperature of 385 K, just above $T_R = 350$ K, the XMCD-PEEM magnetic contrast shown in Fig. 2 reveals clear in-plane ferrimagnetic ordering of Co and Dy elements, both being anti-ferromagnetically aligned with respect to each other in the imaged *x-y* surface plane. Very clearly, the red (blue) regions of the Co XMCD images in Fig. 2 correspond to blue (red) regions of the Dy XMCD signal. Figures 2(a-b) show the magnetic configuration from the antidot array sample and Figs. 2(c-d) from its extended thin film counterpart. Red and blue-color contrast correspond to magnetic domains pointing in the surface plane and projected along the incident x-ray beam direction, with a maximum of ~10% XMCD asymmetry. White (or faint) color contrast indicates either out-of-plane magnetization, magnetization perpendicular to the incident beam direction or the zero net magnetization as it would be expected from the nanohole positions. Comparing the XMCD-PEEM images of the antidot array to the ones of the extended film sample reveals clear differences in the size of the magnetic domains; the antidot sample exhibits nanometer-sized domains separated by white color contrast, whereas the extended thin film sample exhibits domains in the order of several micrometers. This observation suggests that the antidots act as pinning centers for the magnetization stabilizing magnetic nano-domains that are naturally separated from each other. Because the extended thin film lacks of such pinning centers, small nucleation domains which initiate the magnetization reversal have collapsed to form larger ones.

In the following, we present Co XMCD-PEEM images for three selected temperatures from a heating-and-cooling cycle; initial state at 300 K, high-temperature state at 470 K, and the final state reached after cooling back to 300 K again, for both the extended thin film and antidot array samples. Figures 3(a) and 3(d) show the initial states for both samples, where nearly-zero local magnetization signal along the in-plane orientation is observed. The reason for the vanishing magnetic contrast is the out-of-plane spin configuration below $T_R$. Upon heating to 470 K, which is well above the transition temperature, we clearly observe the appearance of in-plane oriented magnetic domains, as shown by the XMCD images in Fig. 3(b) and 3(e). When we cool the samples back to room temperature and re-examine the resulting magnetic properties, the extended thin film sample re-establishes its original out-of-plane spin configuration, as expected [Fig. 3(c)]; however, the antidot array exhibits a non-zero local magnetization as if the magnetic domains are mostly pinned to the in-plane spin orientation even in the absence of any magnetic fields upon cooling [Fig. 3(f)]. To take a closer look at the remanent local magnetization of the antidot array, we extract line profiles along red and blue bits of magnetic domains as well as the real structure as seen by SEM and XPEEM. The results are summarized in Figs. 3(g) and 3(h), where we show line profiles extracted along magnetic images for opposite magnetization directions (blue and red lines). We clearly observe magnetic information repeating every 45-50 nm (between M↑ and M=0 states, and between M↓ and M=0 states), which correlates well with the difference between the center-to-center inter-pore distance and the nanohole diameter. One bit of magnetic information being ~45 nm, one can estimate a data storage density of around 75 Gbits/inch$^2$, which is in the same order of magnitude with the storage density of conventional hard-disk drives.

Figure 4(a) summarizes key experimental results together with a schematic representation of the magnetic anisotropy $K_u$ of the samples as a function of temperature. It shows various channels of writing and storing data using a medium of DyCo$_5$ antidot arrays. The easy axis of the magnetization rotates from out-of-plane to in-plane as the samples are heated above $T_R$, and spins of the two sublattices of Co and Dy align in the sample plane. The change of the easy axis is outlined by the Co XMCD images from an antidot hexagonal unit cell and from the extended thin film sample on the right-hand side of the schematic representation depicting the in-plane anisotropy ($K^{\parallel}$) at high temperatures. Cooling the extended thin film sample back to room temperature leads to either one of the out-of-plane anisotropy configurations in the absence of an external field, as seen in Fig. 3(c). However, the antidot array sample has more options upon cooling: it can decay into two in-plane and two out-of-plane anisotropy configurations, as shown by the Co XMCD-PEEM images on the left-hand side of Fig. 4(a). Regions within the antidot sample with out-of-plane magnetization show a faint color contrast; therefore not much is seen in these images. Two regions of the sample with two hexagonal unit cells of antidots showing in-plane magnetization are also shown. We attribute the occurrence of this magnetic multi-domain state to the presence of nanoholes, which act as pinning centers for the magnetic anisotropy of the sample. To illustrate this in more detail, in Figs. 4(b-d) we show the results of micromagnetic simulations within about one hexagonal unit cell obtained using the OOMMF package [29]. In Fig. 4(b) we show a SEM image from the antidot array superimposed with blue circles around the nanoholes. The circles are used as a mask to construct the input for the calculations. For simplicity, we simulate a Co layer with the magnetic anisotropy oriented along the +*y* direction and an initial out-of-plane magnetization. In Figs. 4(c) and 4(d) we show the calculated magnetization state for the antidot array and the extended film in the fully-relaxed state, respectively. Very clearly, the antidot array in Fig. 4(c) exhibits a magnetic multi-domain configuration, while the extended film remains in a single-domain state in qualitative agreement with the experiments. The calculations clearly reveal that the nanoholes pin the magnetization so that it follows circular contours around them, leading to local changes in the magnetic anisotropy. These changes are caused by the fact that every nanohole gives rise to six constrictions with its first neighbors, so that magnetic domains and domain walls are pinned or trapped by these constrictions. In this way, every domain wall experiences a spatially-dependent dipolar interaction originating from a landscape of pinning potentials within the whole structure, resulting in local rotations of the magnetization and thus in magnetic multi-domain configurations.

Having observed these variety of magnetic domains in the experiment, one can consider controlling their occurrence by applying either in-plane or out-of-plane magnetic field pulses ($H^{\parallel}$ or $H^{\perp}$) to selectively force the system to choose one of the four possibilities mentioned above. The experimental results summarized in Fig. 4(a) are therefore quite analogous to the case where one applies a local magnetic field pulse during the cooling process to manipulate the magnetism of the nano-domains. It would be equally possible to employ this method for both longitudinal and perpendicular recording as they follow the same basic principle. When the energized write head passes by the medium, it leaves behind a magnetization pattern. For perpendicular recording, this magnetization pattern is 'up' and 'down' rather than 'left' and 'right' for longitudinal recording.

Contrary to the high anisotropy HAMR materials (among others: FePt, CoPt, $SmCo_5$), FI antidots such as the ones studied here would be a promising candidate that facilitates writing data by means of manipulating the magnetic easy and hard axes as a function of mild-temperature treatment. The critical temperature of our magnetic system is not the Curie temperature $T_C$ but the spin reorientation temperature $T_R$, which is very close to room temperature. Note that differently from this, high anisotropy HAMR materials need to be heated to temperatures close to their $T_C$, which are around 750-1000 K [30], to sufficiently reduce their high coercivity so that a moderate write field can induce a new magnetic state. In this respect, there have been also attempts to reduce the high $T_C$ of the recording layers with doping. In one example, Thiele *et al.* used Ni doping to reduce the $T_C$ of FePt; however, at the same time the anisotropy decreases [31]. Assuming that in the near future plasmonics and near-field optics will focus and transmit optical energy to spot sizes of around $(25-50 \text{ nm})^2$ and demagnetize the sample [19,30], lower phase-transition temperatures would still be more feasible considering issues such as power consumption, excess-heat dissipation and risk of damaging medium materials.

Summarizing, we fabricated ferrimagnetic $DyCo_5$ antidot arrays for heat-assisted and bit-patterned magnetic recording. Our work demonstrates a novel functionality of ferrimagnetic bit-patterned materials that can be used to write and store magnetic data near room temperature. The size of the nanoholes used in this work was around ~68 nm and their hexagonal lattice constant ~105 nm. We have shown that the magnetization of the antidot array sample can be tuned from an out-of-plane to an in-plane spin configuration upon heating above its spin-reorientation temperature of ~350 K. We have compared the magnetic properties of the patterned sample with the ones of its extended thin film counterpart and observed remarkable differences in the size of the magnetic domains. Our results reveal the formation of small magnetic nano-domains in between antidots with well-defined and stable magnetic domain configurations. Magnetic modulations of the nano-domains contrast are observed to occur within ~45 nm for opposite magnetizations. We have attributed this behavior to the change in the magnetic anisotropy caused by the nonmagnetic inclusions, which at the same time act as pinning centers upon formation of multiple magnetic domains in agreement with our micromagnetic simulations. Finally, we have discussed the implications of our findings for heat-assisted magnetic recording using local magnetic probes. Our work demonstrates that ferrimagnetic antidot arrays exhibit tunable and controllable magnetic traits with the potential to provide superior functionality for energy-assisted bit-patterned media near room temperature.

**FIGURES**

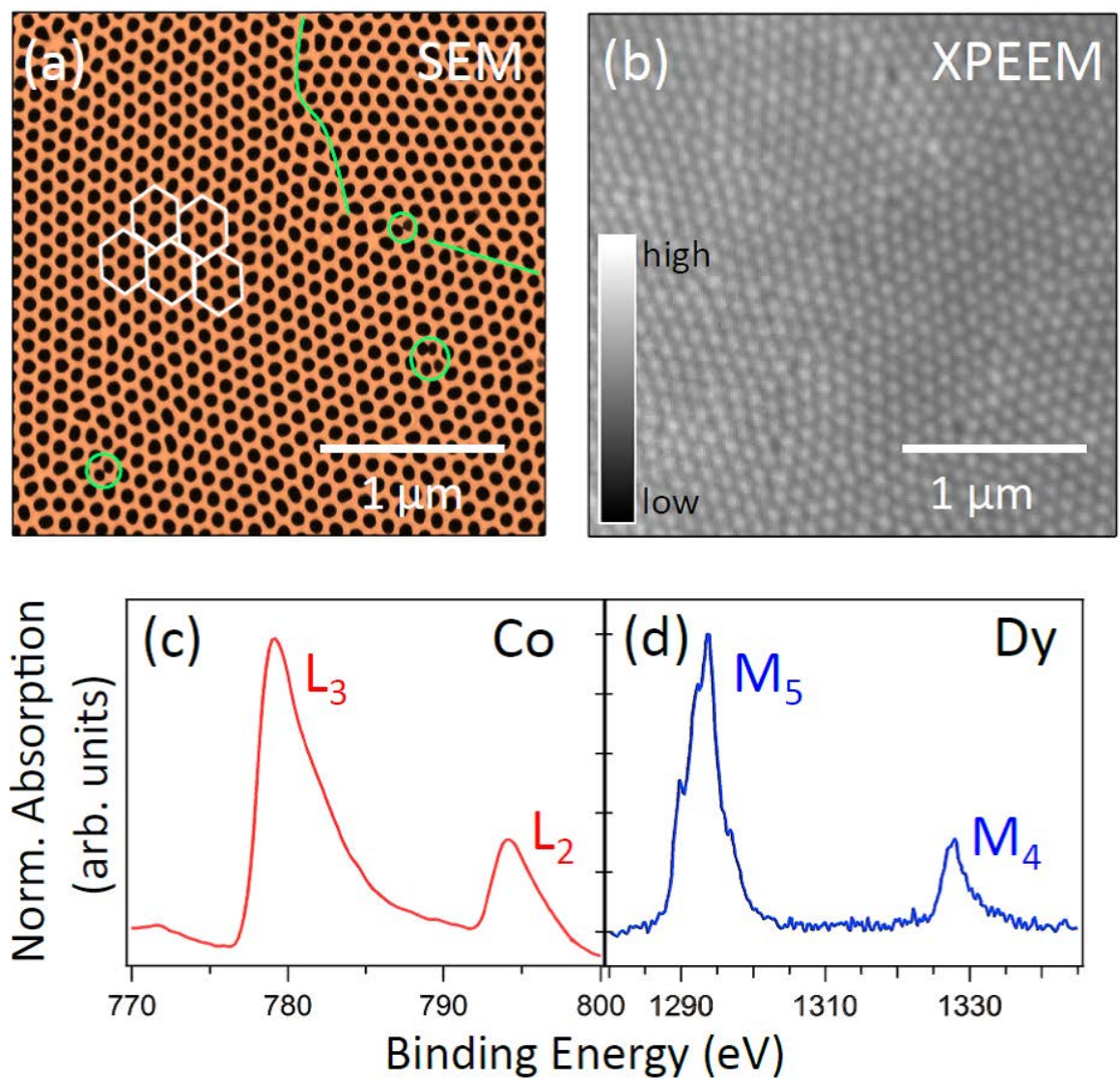

**Figure 1.** (a) SEM and (b) XPEEM images of the hexagonal lattice antidot array of 68 nm pore size and 105 nm separation between pore centers. XAS spectra of (c) Co and (d) Dy elements at their $L_{2,3}$ and $M_{4,5}$ edges, as obtained from XPEEM energy scans using linear horizontal polarized x-rays from the field of view seen in (b).

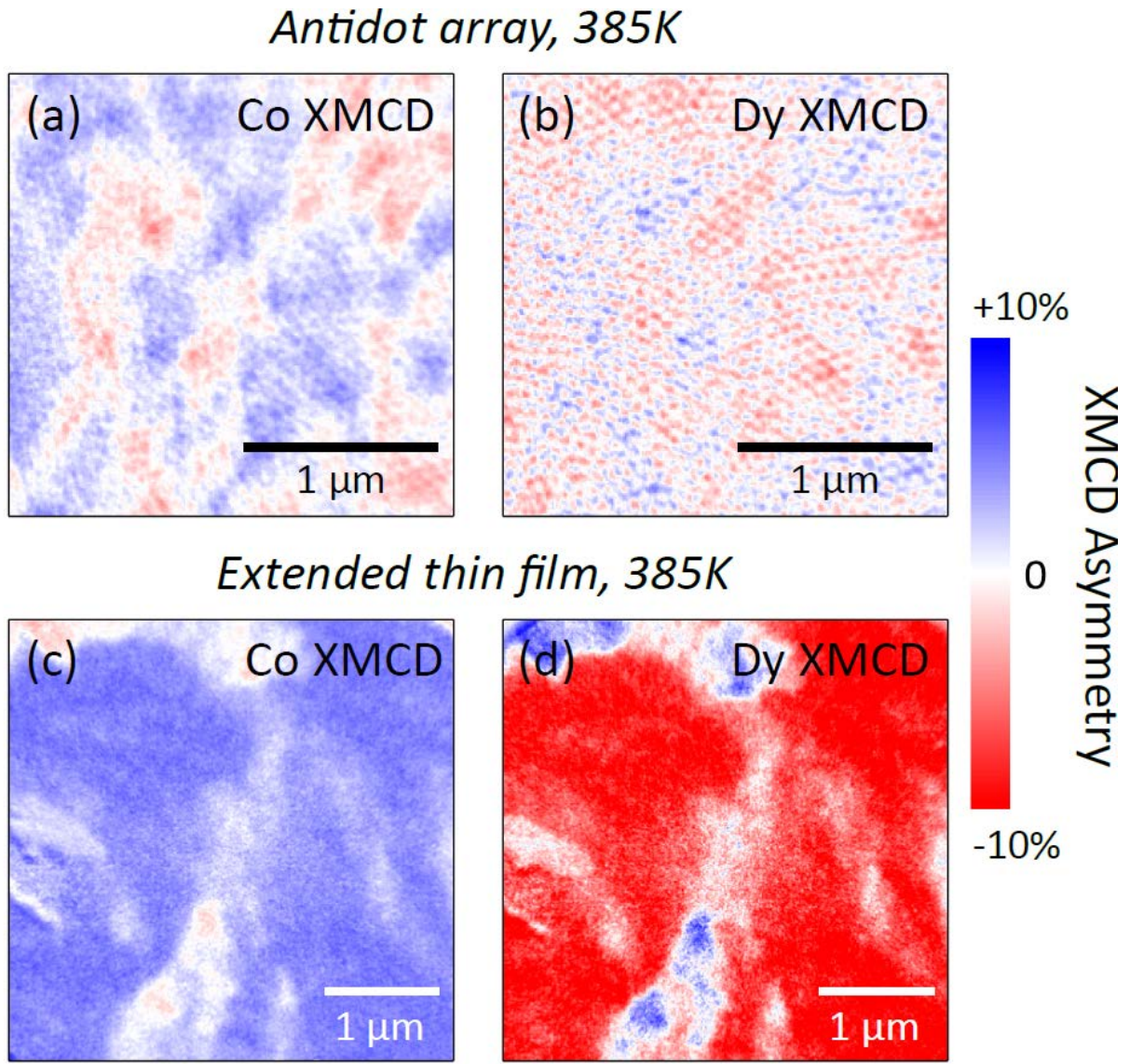

**Figure 2.** XMCD-PEEM images measured at 385 K at the Co $L_3$ and Dy $M_5$ edges from the antidot array (a-b) and the extended thin film sample (c-d) showing the ferrimagnetic ordering of the Co and Dy magnetic moments via the corresponding blue-red XMCD contrast. Antidots stabilize nanometer-sized domains separated from each other.

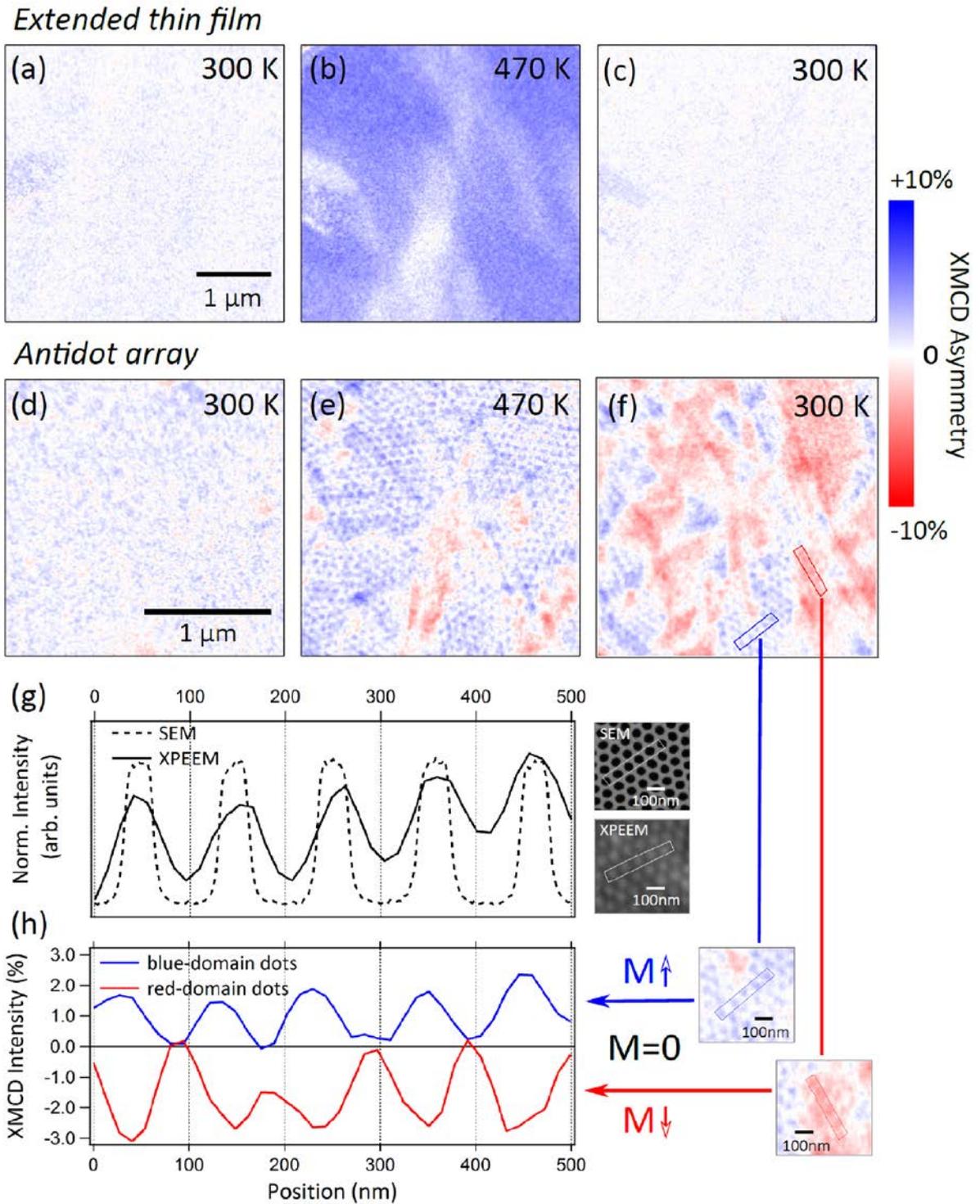

**Figure 3.** XMCD-PEEM images measured at Co $L_3$ edge from the extended thin film sample (a-c) and from the antidot array sample (d-f) at the labeled temperatures. (g) Line profile along the structure of five neighboring antidots, XPEEM (black line) and SEM (dashed line). (h) Line profiles along XPEEM magnetic images for opposite magnetizations (blue and red lines) show magnetic information repeating every 45-50 nm (between M↑ and M=0 states, and between M↓ and M=0 states).

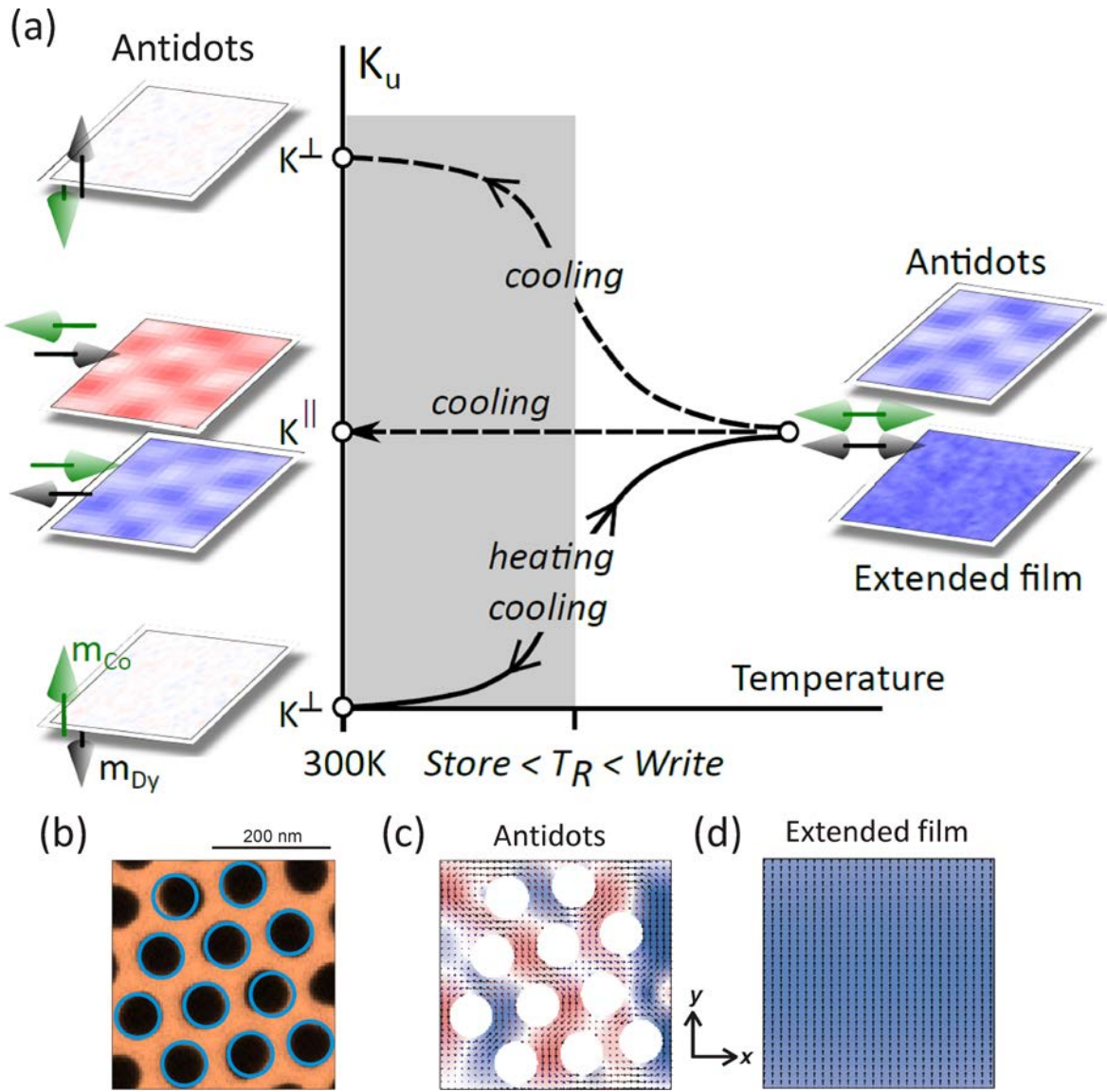

**Figure 4.** (a) Schematic representation of the heat-assisted magnetic recording process and the spin reorientation transition as observed by XPEEM. A selection of images at the Co $L_3$ edge summarizes the experimental results. On the left, the room-temperature results correspond to different XMCD images of the antidot array. (b-d) Results of micromagnetic simulations. In (b), a SEM image from the antidot array superimposed with blue circles around the nanoholes is shown. The circles are used as a mask to construct the input for the calculations. The magnetic anisotropy is oriented along the +y direction and the initial magnetization is out-of-plane. (c) Calculated multi-domain configuration of the antidot array in the relaxed state. Small arrows depict the magnetization directions, blue and red colors the projection of the magnetization along the y direction. (d) Same calculations as in (c) for the extended film, revealing a single-domain configuration.

AUTHOR INFORMATION

**Corresponding Author**

*E-mail: jaime.sanchez-barriga@helmholtz-berlin.de

ACKNOWLEDGMENT

K. J. M and M. V gratefully acknowledge the support from MINECO under project MAT2013-48054-C2-1-R.